# Correction of Wall Adhesion Effects in the Centrifugal Compression of Strong Colloidal Gels.

Richard Buscall[1,2*] and Daniel R Lester

[1] MSACT Research & Consulting, 34 Maritime Court, Haven Road, Exeter, EX2 8GP, UK.
[2] Dept of Chemical & Biomolecular Engineering, The University of Melbourne, VIC 3010, Australia.
[3] School of Civil, Environmental & Chemical Engineering, Royal Melbourne Institute of Technology, Melbourne, VIC 3001, Australia.

*Author to whom correspondence should be addressed at r.buscall@physics.org.

**Abstract**
Several methods for measuring the compressive strength of strong particulate gels are available, including a centrifuge method, whereby the strength as a function of volume-fraction is obtained parametrically from the dependence of equilibrium sediment height upon acceleration. The analysis used conventionally, due to Buscall & White (1987), ignores the possibility that the particulate network might adhere to the walls of the centrifuge tube, even though many, perhaps most, types of cohesive particulate gel can be expected to. The neglect of adhesion can be justified when the ratio of the gel's shear to compressive strength is very small, which it can be well above the gel-point for coagulated systems, although never very near it. The errors arising from neglect of adhesion are investigated theoretically and quantified by synthesising equilibrium sediment height versus acceleration data for various degrees of adhesion and then analysing them in the conventional manner. Approximate correction factors suggested by dimensionless analysis are then tested. The errors introduced by certain other approximations made routinely in order to render the data-inversion practicable are quantified and assessed too. For example, it is shown that the error introduced by treating the acceleration vector as approximately one-dimensional is minuscule for typical centrifuge dimensions, whereas making this assumption renders the data inversion tractable in practice.

## Introduction

Flocculated and coagulated colloidal suspensions form compressible, poroelastic sediments, filter-cakes, drying films and green bodies. The compressive strength of the particulate phase is thus an important property of these systems, which, in the concentrated state tend now to be called attractive or cohesive particulate gels. *Strongly* cohesive particulate gels, by which is meant systems where the interparticle attraction is strong enough to arrest activated processes such as coarsening, creep and delayed collapse, show self-limiting sedimentation and filtration whereby the application of a given acceleration (or load) leads to the formation of a sediment (or filter-cake) with a stable density profile. The scaled well-depth $-U_{min}/k_BT$, the negative of which is taken to be the activation energy for particle escape and diffusion, can be very large for strongly-flocculated and coagulated colloidal particles, > 100, hence it is found that voluminous sediments and cakes can be stable for years or even decades, this being consistent with the approximately exponential dependence of the Kramers' time upon well-depth. Under such circumstances it is meaningful operationally to speak of equilibrium sediment heights and filter cakes thicknesses and to attempt to calculate the curve of compressive strength as a function particulate volume-fraction (φ) from curves of equilibrium filter-cake thickness (or sediment thickness) versus load (or centrifugal acceleration).

Several methods have been used to determine the compressive strength as a function of density. The most direct in concept is pressure-filtration, whereby the equilibrium filter cake volume is measured for a series of applied pressures in carefully-controlled and instrumented pressure-filters [1-8]. The analysis of this test is trivial, or, it is provided that the possibility or effect of adhesion of the particulate network to the side-walls of the filter can be ignored. It is to be expected from the pioneering work of Michaels and Bolger [9], that such effects can only safely be neglected when the quantity $\frac{4h\tau_w}{PD} \ll 1$, where $\tau_w$ is the adhesive strength at the walls, *P* is the compressive strength, and *h* and *D* are the thickness and diameter of the cake respectively. $\tau_w$ here is taken to be the critical or yield stress required to cause debonding and slip at the walls and hence it can be determined independently in principle by using, say, a suitable rheometer fitted with smooth tools constructed of material similar to that of the walls of the pressure chamber. Alternatively, one might simply use the true yield stress of the suspension, $\tau_y$, as an upper estimate of the adhesive strength, since it has been found in practice that whereas $\tau_w < \tau_y$, it is generally of similar order, with values of $\tau_w/\tau_y$ in the range 0.2 to 0.6 being found for coagulated systems [8,10-13]. Another option, perhaps, would be to use several different filter diameters and extrapolate the results to infinite diameter, although this might not be too convenient in practice since it requires a set of several pressure filtration cells and a rather large amount of sample overall.

The other three methods commonly used rely on sedimentation-equilibrium in one way or another:

(i) Measure the equilibrium sediment height in a series of batch settling columns of increasing height under normal gravity [1,8,14-16].
(ii) Measure the equilibrium height in a suitable laboratory centrifuge for a series of accelerations [1,8,10,16-22]. The centrifuge should be fitted with a swing-out rotor, preferably, since otherwise the total resultant acceleration is not in the plane of the tube axis except asymptotically at high 'g'.
(iii) Determine the equilibrium concentration versus depth profile under gravity in a column tall enough to produce significant compression towards the bottom [8,23-29,30-32].

The principle behind all of these methods is that the unbuoyed self-weight, $w(h) = \Delta\rho g \int_h^{h_{eq}} \phi(z)dz$, at each point $h$ in a sediment of total equilibrium height $h_{eq}$, is balanced by the compressive strength $P(\varphi\{h\})$: here $\Delta\rho$ is the suspension inter-phase density difference, $g$, the magnitude of the gravitational or centrifugal acceleration and $\varphi(h)$, the local volume-fraction. The balance above only holds exactly in the absence of adhesion, whereas the vertical force balance [9,33].

$$\frac{dP}{dz} = \Delta\rho g \phi - \frac{4\tau_w}{D}, \tag{1}$$

suffices to show that wall adhesion effects should only be neglected when

$$\frac{4\tau_w[\phi(0)]}{D\Delta\rho g \phi(0)} << 1 \tag{2}$$

Methods (i) and (ii) are very similar in basis, clearly, the prime reason for distinguishing between them being that adhesion affects them very differently, in part because the acceleration $g$ is much larger in centrifugation. A second reason is that, whereas the raw data in case (i) of $h_{eq}$ versus $h_0$ can be inverted exactly to obtain $P(\varphi)$ when adhesion is negligible, one of the corresponding pair of equations used to invert centrifuge data, that for the pressure, is only approximate. The particular approximation used conventionally is only good to within ca. 5% in the pressure, typically [18]. It is however possible to do better and invert the data from the centrifuge method significantly more accurately, as will be shown below.

Like pressure filtration, method (iii), concentration-profiling, is fairly direct in the absence of adhesion, since integration of the concentration profile gives the pressure at any height $z$ above the base of the column, such that,

$$\Delta \rho g \int_z^{h_{eq}} \phi(z') dz' \rightarrow P(\phi[z]) . \qquad (3$$

It has however been shown recently [27,28] that the method is rather sensitive to any adhesion to the walls of the tube, to the point where it is possible to invert raw data obtained from two or more columns of different diameter [27] so as to determine both the compressive and the shear strength functions fairly accurately. The exact method of inversion described in [27] is way too computationally intensive to be used routinely, it has however subsequently proved possible to develop a simple but accurate analytical method based upon an approximate pseudo-1-dimensional force balance [33].

The essential problem with the profiling method is that any errors coming from adhesion accumulate down the column [27]. This can be seen by reference to eqn 2 as since the shear yield strength $\tau_w$ in the numerator must increase with $\varphi$ at least as fast as $\varphi^2$, the ratio on the LHS must increase with $\varphi$ and thus depth. This means that in a tall column of modest diameter eqn 2 will always be violated below some critical height, causing the apparent compressive strength to appear to diverge at the corresponding $\varphi$ if the wall effect is ignored in the analysis. Thus, the apparent or uncorrected compressive strength obtained using narrow columns can appear to diverge well below the jamming limit when adhesion is significant in this sense [27]. It should be noted that the the tubes need not be all that narrow in laboratory terms as in ref. 27 it was necessary to go to tube diameters of > 100mm in order to render the wall effect small. The basic problem with the profiling method, then, is that in order to determine $P(\varphi)$ over a significant range of concentration, the column needs to be tall in order to obtain substantial compression near the bottom, except that this tends to equate to working at large 4 $h_{eq}/D$ which just amplifies errors coming from adhesion. Method (i), whereby the sediment is compressed more and more by increasing $h_0$ can suffer badly from this problem. By contrast, method (ii), centrifugation, involves increasing the acceleration at fixed $h_0$ and for this and other reasons wall effects are of less consequence, as will be seen. In a centrifuge the pressure is increased by increasing the rotation rate $\omega$, and thus acceleration, hence any errors coming from the neglect of adhesion now decrease going up the $P(\varphi)$ curve, everything else being equal, simply because the sediment is shrinking, causing the area in contact with the wall reduce. The same is true for pressure-filtration. The other significant advantage of the centrifuge method is its large dynamic range, since four or even five decades of acceleration can be obtained using fairly standard laboratory centrifuges (the range $10^2 < g = \omega^2 R < 10^6$ ms$^{-2}$ is accessible using standard centrifuges fitted with swing-out rotors; here *R* distance from the axis of rotation to the bottom of the centrifuge tube). Its main disadvantage is that it can be difficult to get much below $10^2$ ms$^{-2}$ and hence close to the gel-point with most standard laboratory centrifuges, since control of speed can be problematic at the bottom end of their notional speed range. It is however possible to custom build or adapt suitable centrifuges of course. It can likewise hard to get close to the gel-point in pressure-filtration because of the problem of applying and controlling small pressures.

A word needs to be said about adhesion versus cohesion perhaps. There are ca. ten different types of interparticle attraction and hence drivers of aggregation known (e.g Van der Waals, depletion, liquid-bridging, incipient flocculation, charge-patch, bridging, and so on [30]). Many of these are indiscriminate inasmuch that they can or will produce an attraction between particles and any bounding surface too, even if the strength of attraction might depend upon the nature of the material in some cases (cf. Van der Waals), if not others (e.g. depletion). The Derjaguin approximation [30] suggests that the individual particle-wall force should be ca. *twice* the particle-particle force for spheres, everything else being equal. This alone could be taken to imply that the adhesive shear strength could well be higher than the bulk shear strength, except this has never been observed in experimentally to the best of our knowledge [8,10-13]. Both of the two shear strengths can be measured rheometrically in principle, by determining the shear yield stress twice, first, say, using smooth shearing surfaces, to obtain the wall yield stress, and then again with suitably roughened tools, so as to obtain the true or bulk yield stress[13]. An alternative would be to use a vane tool and two smooth outer cylinders, one to give a narrow gap, such that yield occurs prematurely at the outer wall, and the second to give a wide gap such that yield then occurs at the vane [13]. Such little data as there is in the literature suggests that $0.2 \leq \tau_w/\tau_y \leq 0.7$ is typical, supposing that the limited published data is reasonably representative [8,10-13].

**Constitutive relationships and modelling**

Quantitative constituitive relationships were needed in order to generate synthetic sedimentation-equilibrium data for subsequent analysis. More specifically, the continuum model described in detail by Lester et al. [27] requires the compressive yield stress and wall shear functions $P_y(\varphi)$ and $\tau_w(\varphi)$ as input. Following Lester et al. [27], the functional form used for $P_y(\varphi)$ was,

$$P(\phi) = k\left[\left(\frac{\phi}{\phi_g}\right)^n - 1\right]; \quad \phi > \phi_g \tag{4}$$

This form captures the experimentally-observed rapid increase in compressive strength at solids concentrations above gel-point, along with the progression towards power-law behaviour seen typically at higher concentrations. The power-law index $n$ has been found to be ca. 4 typically for electrolyte-flocculated systems [8,10, 16-22, 33] with strongly polymer-flocculated $CaCO_3$, as used in Lester et al.[27] showing a somewhat higher value of ca. 5. In consequence, numerical data was synthesised for both $n = 4$ and $n = 5$. Please note, by the way, that eqn 4 neglects the expected divergence of the compressive strength at jamming, although this feature can of course be built into eqn 4, if needs be.

Again, following Lester et al. [27] and guided by experimental data [11], it was assumed that $\tau_w(\phi) \leq \tau_y(\phi) = \gamma_c G(\phi)$ and, in turn, that the linear shear modulus $G(\varphi) = 5/3 \, K(\varphi) = 5/3 \, dP(\varphi)/d\ln \varphi$, where $K$ is the bulk modulus of the particulate network. It should be emphasised that, here, $\gamma_c$, the apparent yield strain is merely a parameter; an apparent critical strain *defined* by the eqn $\gamma_c \equiv \tau_y / G$. The only reason for parameterising thus is that the apparent yield strain, so defined, varies much more weakly with volume-fraction than does either $\tau_y$ or $G$ [8,10-12,19-21,34], making it a convenient way of quantifying the intrinsic adhesive strength of a system. Hence, in the simulations, the amount of wall adhesion could be varied conveniently by changing the value of the apparent critical shear strain $\gamma_c$, defined as above, The latter has been found to vary widely in practice from one material to another. For electrolyte aggregated systems values from 0.00005 to 0.025 have been reported [8,10-12,19-21, 34] with, 0.0001 to 0.002 being more typical perhaps. Larger apparent critical strains of ca. 0.02 have been found for strongly polymer flocculated suspensions and a value of ~ 0.01 has been found enough to give strong wall effects in gravity-settling even in quite large diameter tubes [27]. From considerations of the nature of colloidal inter-particle forces the critical strain might be expected to vary inversely with particle size, and weakly, with volume-fraction, viz. to first order or less. Published data are reasonably consistent with these notions, so far as the data go [8,10,11,21, 34]. The functional forms used for the compressive strength combine to cause the ratio of the two, $\tau_w(\phi) / P_y(\phi)$ to decay rapidly from unity at the gel-point to an asymptotic value of $\tfrac{5}{3}\gamma_c$ at high volume-fraction, as shown in fig.1. The rate of decrease is such that the ratio of shear to compressive stress is approximately constant over much of the range, which is what has often been seen experimentally too[8,10,11,21, 34]. On the other hand, the ratio has always to approach unity at the gel-point from theoretical considerations[11,27,29,31,33,34], and wall-adhesive effects then to be expected have been observed experimentally near the gel-point [27,28]. Indeed, the constitutive forms used here are exactly those that were used earlier to fit experimental data for $CaCO_3$ particles flocculated using three different polymeric flocculants [27, 33].

Synthetic sedimentation curves were generated using the functional form (4) for apparent critical strains $\gamma_c$ of 0, 0.0002, 0.002 and 0.02. The gel-point and starting volume-fractions were taken to be 0.1, the initial column height $h_0$ was 0.075m and the tube diameter was taken to be 10m in most of the runs, latter dimensions being fairly typical of centrifuge testing.

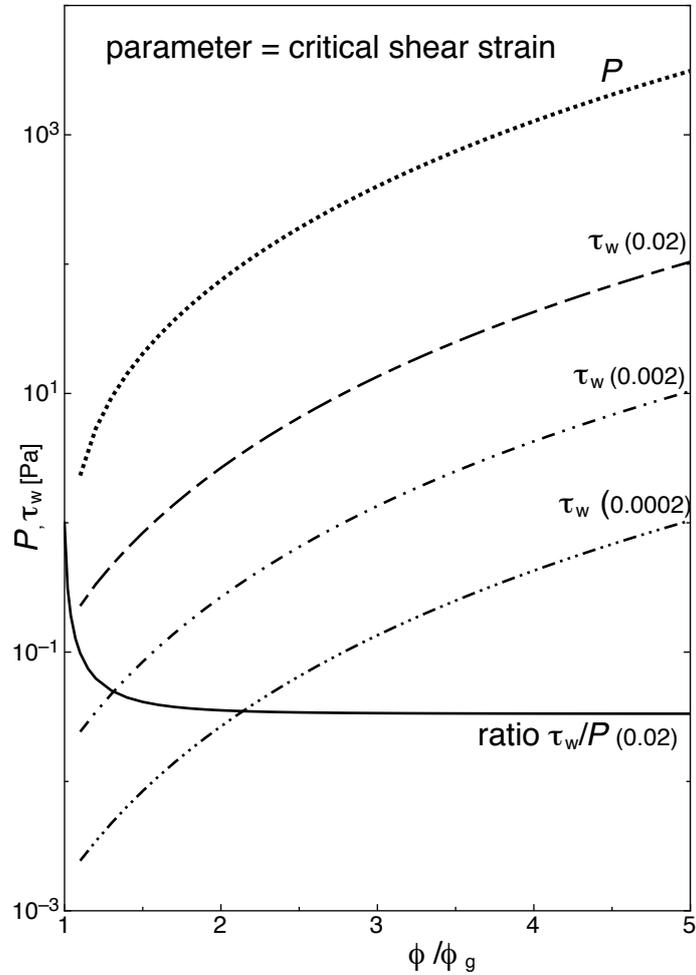

**Fig. 1:** Input material functions: compressive strength and wall stress, the latter for three values of the apparent critical shear strain. The ratio of wall stress to compressive stress is plotted for the largest strain. It can be seen that most of the variation is confined to a region close to the gel-point.

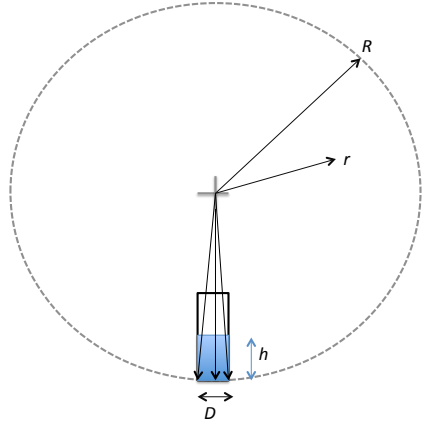

**Fig. 2:** Schematic illustration of the geometry of the centrifuge test.

In any real centrifuge the magnitude of the acceleration $g = \omega^2 r$ varies with radial distance away from the axis of rotation and hence it also varies down the column of suspension. It is useful to define maximum and mean values of the magnitude of the acceleration thus, (cf. Fig. 2),

$$g_{max} = \omega^2 R; \quad \bar{g} = g_{mean} = \omega^2(R - h_{eq}/2), \qquad (5$$

where the spatial mean can be thought of as representing the mean acceleration per unit mass that would be experienced by a perfectly incompressible suspension at equilibrium, whereas the maximum acceleration is that for a ficticious material having zero rigidity. It should be clear then, that the mass average acceleration in any real equilibrium sediment must bounded by the maximum and mean accelerations in (5).

The task of developing and validating simple but robust means of data inversion that acknowledge wall adhesion can be largely carried out in the uniform $g$ limit, since the case of spatially-varying acceleration can be handled in the same way, as will be shown below.

In the base case of both uniform $g$ *and* negligible wall adhesion, the equilibrium curve of $h$ versus $g$ can be inverted exactly using the simple formulae [18],

$$P_y(\phi_b) = \Delta \rho g h_0 \phi_0; \quad \phi_b = \frac{h_0 \phi_0}{h_{eq} + dh_{eq}/d\ln g} \qquad (6$$

to give the pressure and the volume-fraction at the bottom of the tube. It can be seen on the RHS of (6) that, in this ideal case, the compressive strength itself depends parametrically upon $g$ plus three known constants only, hence any scatter in the raw data of $h_{eq}(g)$ and any errors arising from numerical or graphical differentiation affect only the volume-fraction.

In the case of radially varying $g$ the data can only be inverted approximately even when adhesion is absent or ignored [18]. The approximation suggested by Buscall & White [18] and used by most workers to date reads,

$$P_y(\phi_b) = \Delta \rho \overline{g} \phi_0 h_0$$

$$\phi_b = \frac{\phi_0 h_0 \left[1 - \frac{1}{2R}\left(h_{eq} + \frac{dh_{eq}}{d\ln g_{max}}\right)\right]}{\left[\left(h_{eq} + \frac{dh_{eq}}{d\ln g_{max}}\right)\left(1 - \frac{h_{eq}}{R}\right) + \frac{h_{eq}^2}{2R}\right]}$$

(7

It can be seen that the eqn for the pressure is now approximate, since it contains the mean acceleration. Eqns 7 were derived by acknowledging the radial increase of $g$ down the tube axis, but by supposing otherwise that the **g** vector field within the tube is parallel to the column walls, and constant at any height $h$. This "parallel-g" approximation, valid for $D/2\pi R \ll 1$, simplifies the analysis of the test very considerably by rendering the analysis 1-dimensional. $D/2\pi R$ is of order 0.01 in a typical benchtop centrifuge and so the error introduced by the use of the parallel-g approximation is be expected from the geometry to be of order $1 - \cos(D/R) \sim (D/R)^2$. It will be shown later via a direct comparison of simulated data for parallel and true vector **g**, that the error is indeed negligible to all intents and purposes for small $D/2\pi R$. Furthermore, if both $D/2\pi R \ll 1$ *and* $h/R \ll 1$, then the acceleration can be regarded as both parallel and approximately uniform (as per eqn 5), allowing eqns 6 to be applied to centrifuge data, and the validity of this approximation for small but finite $(D/2\pi R)$ will be tested also.

The numerical calculations required to generate synthetic sedimentation-equilibrium curves in centrifugation, i.e. $(P_y(\phi), \tau_w(\phi)) \to (h_{eq}, \overline{g})$, can be executed efficiently via plasticity theory, although the usual problems associated with spurious slip-lines are encountered [27]. Because of this, it is preferable to resolve the deviatoric stresses below the yield stress by incorporating solid-like viscoelasticity sub-yield. For materials that are brittle in shear ( have a small critical shear strain $\gamma_c$ in the current terms), and for the purposes of calculating sedimentation equilibrium, the results are totally insensitive to the precise behaviour assumed and so small-strain linear elasticity suffices to regularize the solutions given by plasticity theory, as has been shown and discussed at length elsewhere [27]. The reader is referred to that work [27] for details of the method employed to simulate the equilibrium sedimentation curves. The values of the various constants and variables used are shown in table 1 below.

**Table 1: Values of key quantities used in the simulations**

| Quantity | Symbol | value |
|---|---|---|
| Initial vol. fraction | $\varphi_0 = \varphi_g$ | 0.10 |
| Compressive strength Pre-factor | $k$ | 5 Pa |
| Compressive strength exponent | $n$ | 4, 5 |
| Interphase density difference | $\Delta\rho$ | 2000 kg m$^{-3}$ |
| Tube diameter | $D$ | 5, 10, 15 mm |
| Initial column height | $h_0$ | 0.075m |
| Centrifuge radius | $R$ | 0.175m |

**True radial versus parallel versus uniform acceleration.**

Simulated equilibrium height versus acceleration curves calculated for true vectorial **g** and constant parallel-g are plotted in fig. 3 for $D/2\pi R$ = 0.009 (D=10mm) and $h/R \sim$ 1/3 and a power-law index $n$= 4.0. It can be seen that even for $h/R$ as large as 1/3, the radial variation of acceleration makes little difference to the predicted equilibrium curve; provided, that is, that the heights from the vector simulation are plotted against the mean magnitude of the acceleration and not its maximum value. This implies that for $\frac{h}{R} \leq \frac{1}{3}$ the simpler eqn (6) can be used to invert real centrifuge data rather than (7), even though both are equally easy to use. The main difficulty in the application of either of these equations to experimental data is the calculation of the slope d$h_{eq}$/dln $g$ which needs to be evaluated as accurately as possible. Fortunately, empirical observations suggest that $h_{eq}$ as a function of ln(g) has only weak curvature (arising from power-law behaviour of $P_y(\varphi)$ away from the gel-point), which facilitates 1$^{st}$ order piece-wise fitting for the purpose of smoothing and differentiating real experimental data. Since uniform $g$ and true vector **g** differ only slightly, then parallel **g** and true vector **g** must only differ by an even smaller amount, as is confirmed in Fig.

4. The error introduced by the parallel-**g** approximation is negligible, clearly, as expected from scaling, $\sim (D/R)^2$, in spite of previous criticisms regarding its use [35, 36]. Those criticism not withstanding, and geometrical scaling estimates apart, it is abundantly clear from the simulated data in fig. 4 that the parallel-**g** assumption is a very accurate approximation for real laboratory centrifuges, and hence a negligible source of error.

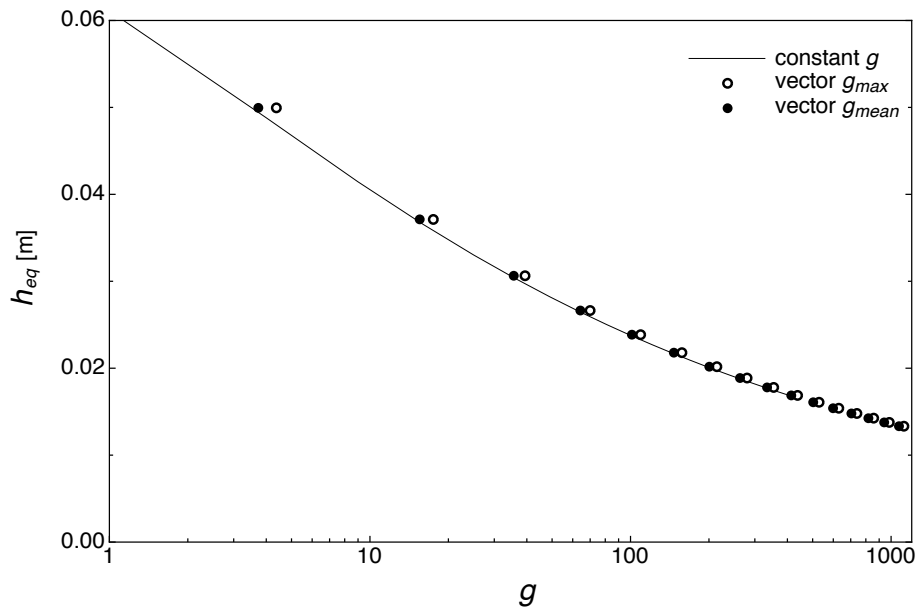

**Fig. 3** Synthesised equilibrium height versus acceleration data for constant $g$ compared with centrifugal acceleration at two equivalent mean accelerations of $g_{mean} = w^2(R-h/2)$ and $g_{max} = w^2 R$ for $n=4$, $D=15$mm and $h_{eq}/R < 1/3$.

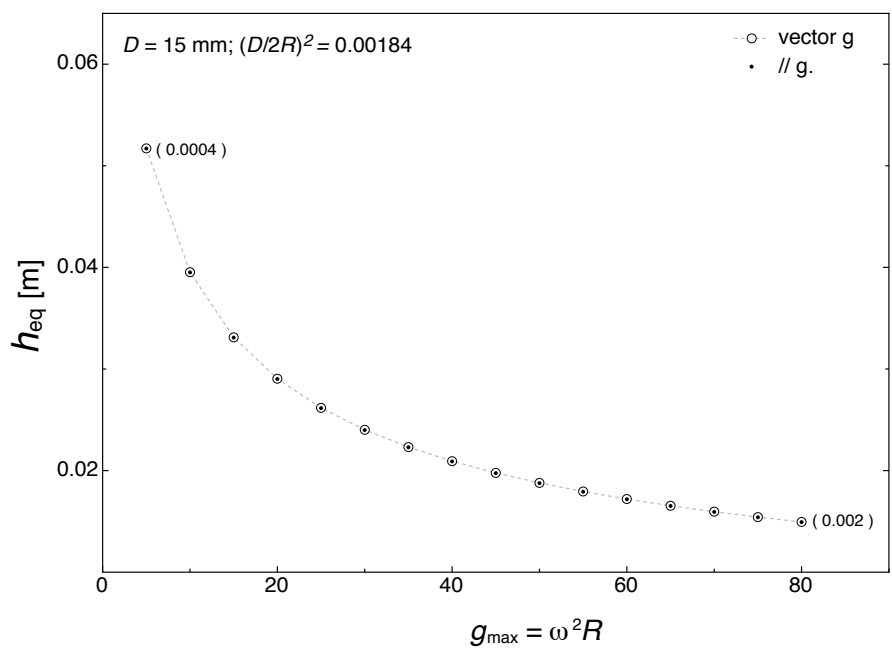

**Fig. 4** Synthesised equilibrium height versus acceleration data for vector centrifugal *g* compared with the parallel-*g* for 15mm internal dia. tubes and *n*= 4. The figures in brackets show the fractional error of the parallel-g approximation at each end of the curve.

**Effect of adhesion**

A set of synthetic equilibrium height data for both *n* =5 and *n*=4 is shown in fig. 5. While the effect of increasing the critical strain from 0.0002 to 0.002 is barely discernible, that of going from 0.002 to 0.02 is more significant. For the suspension constitutive model used here the corresponding yield stress ratio, $\tau_w(\phi)/P_y(\phi)$, decreases rapidly from ca. unity at the gel-point toward the asymptotic value of $5/3\,\gamma_c$ at high concentrations. It should be emphasised though that a small shear to compressive yield stress ratio does not of itself necessarily always equate to a small effect of adhesion though, since wall effects are amplified by a geometric factor of ~ $4h_{eq}$ /D, hence they will always appear if the tubes are too narrow [9, 27].

The synthetic data were inverted using (6), and some sample output is plotted in fig. 6. It can be seen that the large error introduced by neglect or ignorance of adhesion near the gel-point decreases with increasing volume-fraction. This decrease has two components, the first of which is due to the fact that the surface area of the gel in contact with the walls decreases as the gel compresses, the second of which comes from the constitutive behaviour assumed, whereby the wall-adhesion to compressive strength ratio decreasing rapidly with volume-fraction (cf. fig.1): this type of behaviour is consistent with various experiments [11, loc. cit, 27] and consistent too with the notion that, whereas cohesive particulate networks are short in shear, they exhibit ratchet poroelastic and strain-hardening in compression [11,31-34]. Similar results for *n*= 4 are shown in fig. 7. The same pattern can be seen, viz. large errors due to the neglect of adhesion for a critical strain of 0.02 near the gel-point, but with convergence towards the correct result at high concentration.

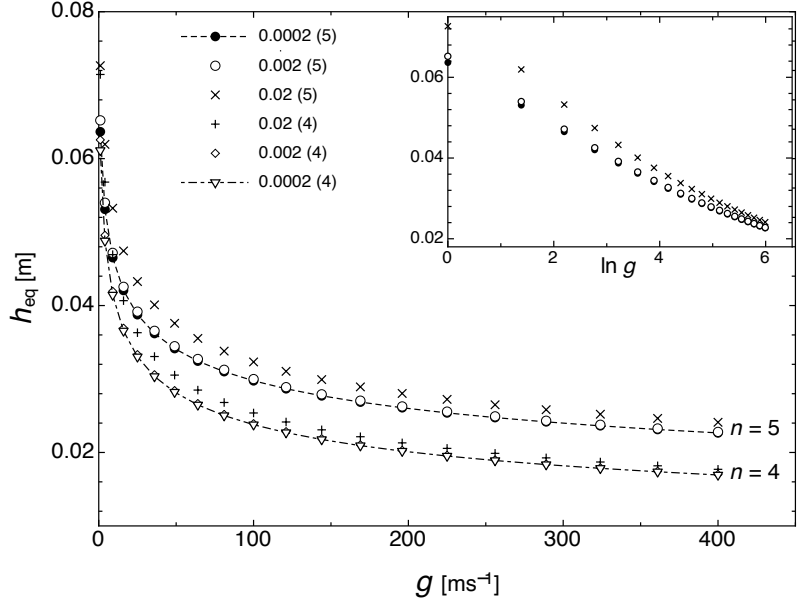

**Fig. 5** Synthesised equilibrium height versus acceleration data for constant $g$, 10mm internal diameter tubes, two values of the power-law index and for three different levels of wall adhesion as parameterised by the critical shear strain. The effect of adhesion is just discernible at 0.002 and significant at 0.02. The result for zero shear strain (not shown) is indistinguishable from the curve for 0.0002. The inset semi-log plot illustrates the slow change in the gradient term needed in eqns (7) and (6).

Since rheological characterisation will often involve determination of the either or both of the adhesive or true shear yield stresses, it is pertinent to ask whether the apparent compressive strength determined by applying eqns (6) or (7) be can be corrected using a set of shear yield stress data measured separately using some sort of rheometer. An overall momentum balance of the type first described by Michaels and Bolger [9] (cf. eqn 1) motivates the testing of possible corrections of the form,

$$P_{corr.}(\phi) = P_{app}(\phi) - 4\tau_w(\phi)\frac{h_0}{D}f(h_{eq}) \qquad (8$$

wherein $f(h_{eq})$ represents an approximate or average way of accounting for the effect of the decreased volume and thus surface areas of the sediment on the total adhesion at each equilibrium point. Thus, by reference to (8), the following simple forms are suggested as likely over-corrections,

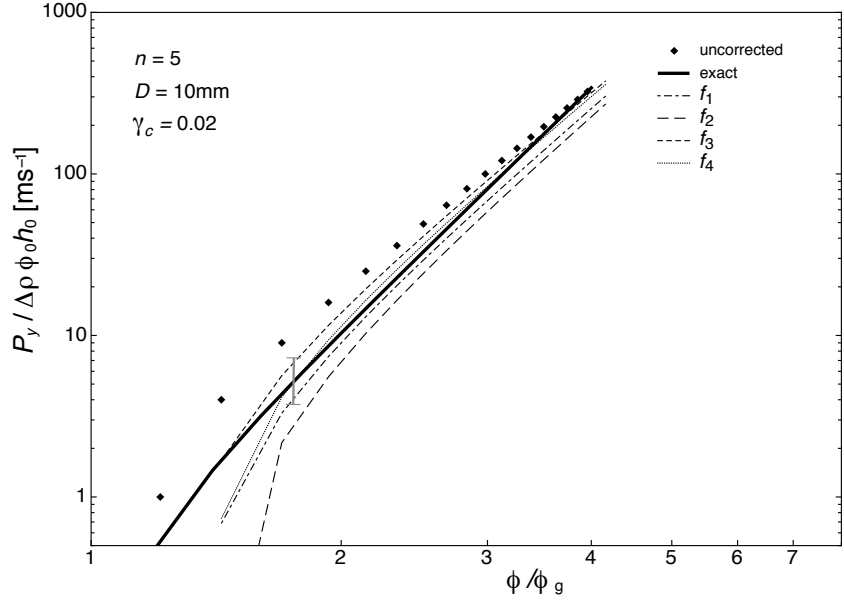

**Fig. 6** Apparent compressive strength for critical adhesive shear strain = 0.02 obtained by applying eqns 3 to the data in fig. 5 (points) compared with the true curve (heavy line). Also shown are the results of applying the suggested under and over corrections $f_1$ to $f_4$ to the points. Approximate corrections of this type cause the result to become negative below ca. 1.3 times the gel-point. Below that only iteration of the data via retrodiction and correction will do and even that scheme will fail if the gel is supported entirely by the walls.

$$f_1(h_{eq}) = \frac{h_{eq}}{h_0} \quad \text{or,} \quad f_2(h_{eq}) = \frac{\phi_0}{\phi} \quad , \tag{9}$$

whereas the following pair are likely to be under-corrections

$$f_3(h_{eq}) = \left(\frac{h_{eq}}{h_0}\right)^2 \quad \text{or,} \quad f_4(h_{eq}) = \left(\frac{\phi_0}{\phi}\right)^2 \quad . \tag{10}$$

That this is so should be fairly obvious, since the first pair become exact in the limit of infinite strength (or, zero compressibility) and the second in the opposite limit of zero strength (or, infinite compressibility). It should also be mentioned though that we have, nevertheless, put a substantial amount of effort into attempting to derive such a set of corrections theoretically by trying to find a general but accurate pseudo-1d approximation scheme analogous to that developed earlier for uniform g [33]. This goal has however defeated us to date, since the problem becomes very formidable even for 'parallel-g'.

The candidate upper and lower bound corrections (9) and (10), applied to the data for critical strain of 0.02, are compared with the true curve in figs. 6 and 7. The corrected curves bracket the true curve except near the gel-point, where no simple correction term is ever going to eliminate the errors coming from unacknowledged or unsuspected adhesion, since it *always* dominates very near the gel-point. The difficult region looks to be $\phi/\phi_g \leq 1.3$, where iteration may be required order to recover the true curve, either that or simply using wider tubes: the discriminator though is the factor $\dfrac{4\tau_w h_0}{P_{y,\text{app.}} D}$. If the latter is significantly less than unity, then the lower-bound corrections $f_3$ or $f_4$ should be good, as the errors are then always fractional.

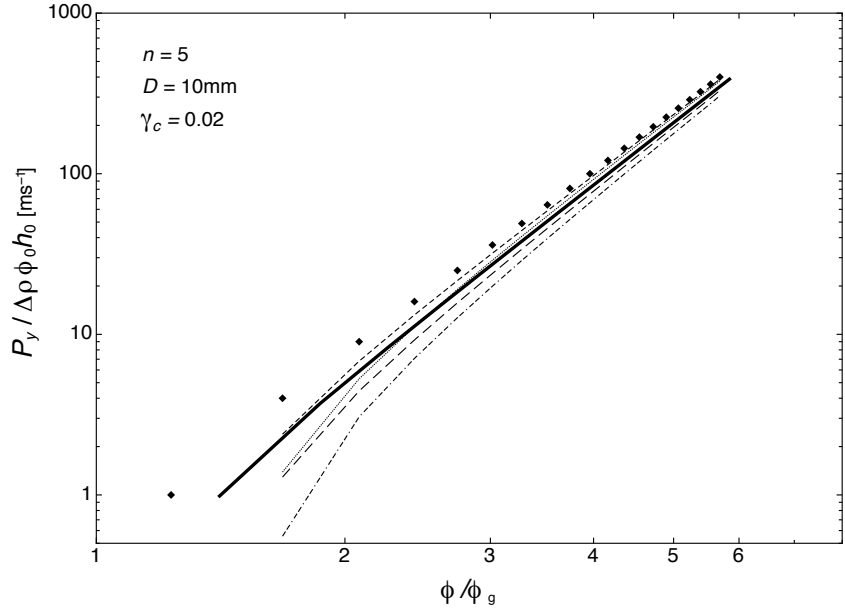

**Fig. 7** As for fig. 6 but for a power-law index of *n* = 4. Reducing the power-law from 5 to 4 reduces the errors coming from adhesion a little, but otherwise the picture is much the same.

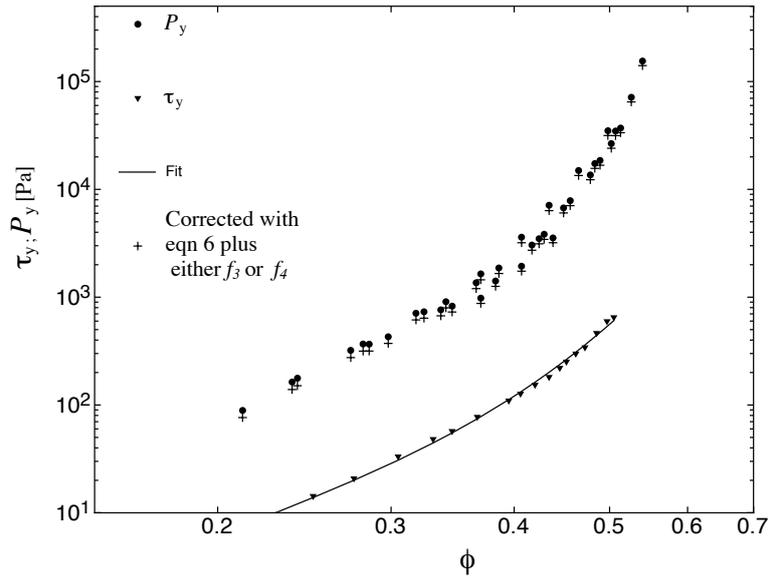

**Fig. 8** Data of Zhou et al. (replotted from ref. 10, see also ref. 8, fig. 3) and corrected according to eqn 6 using correction $f_3$ from eqn10 and the fit to the shear data shown. Note that correction $f_4$ gives similar results.

A data set taken from the literature[10] is re-plotted in fig. 8, together with the lower-bound corrections calculated from the shear yield stress data shown. The latter were measured using a cruciform cross-sectioned vane and hence they are true yield stress values and hence over-estimates of the adhesive or wall stress needed. This almost certainly means that the corrections are systematically too large, not that it would seem to matter too much in this case as they rather small anyway.

What if one does not have a means of generating independent shear-stress data? Or, that one does not wish to undertake the fairly substantial workload involved; the centrifuge generates the complete curve, whereas for the shear rheometry one needs a whole series of samples spanning the concentration range and to be sure that the structure is preparation or history independent. It is fairly obvious that an alternative would be to vary the column height and the tube diameter and extrapolate each point on the sedimentation curve to $h_0/D \to 0$. It should be possible in principle to determine both strength functions by this means too, although the analysis could well be rather tedious, perhaps. The real problem with this method though is that one can only realistically vary $h_0/D$ by a factor of about 5 even in the most accommodating of centrifuges, whereas a decade or more would probably be needed in order to extract reliable values for the shear strength. On the other hand, varying $h_0/D$ by just two or three can and has been used to confirm that adhesion is negligible for various materials, including coagulated latex [16-20].

The simulated effect of diameter is illustrated in figs. 9 and 10 for the most adhesive case studied here, i.e. critical strain = 0.02, and for three different values of tube diameter at fixed $h_0$ = 0.075m. The tube diameters were 5, 10, 15mm, but also shown in fig. 10 is the datum for infinite diameter at each acceleration. The effect of diameter decreases with increasing acceleration as expected and, for the two higher accelerations and hence well away from the gel-point, the extrapolation to $1/D = 0$ is sensibly linear, whereas at g =5 it is curved away from the abscissa somewhat, implying that the use of three narrow centrifuge tubes might not be quite good enough for more adhesive cases nearer the gel-point. It is always possible though to augment centrifuge data with gravity-settling data of course, where it is much easier to cover a wide a range of tube height to diameter ratios $h_0 / D$, supposing that material availability is not a problem.

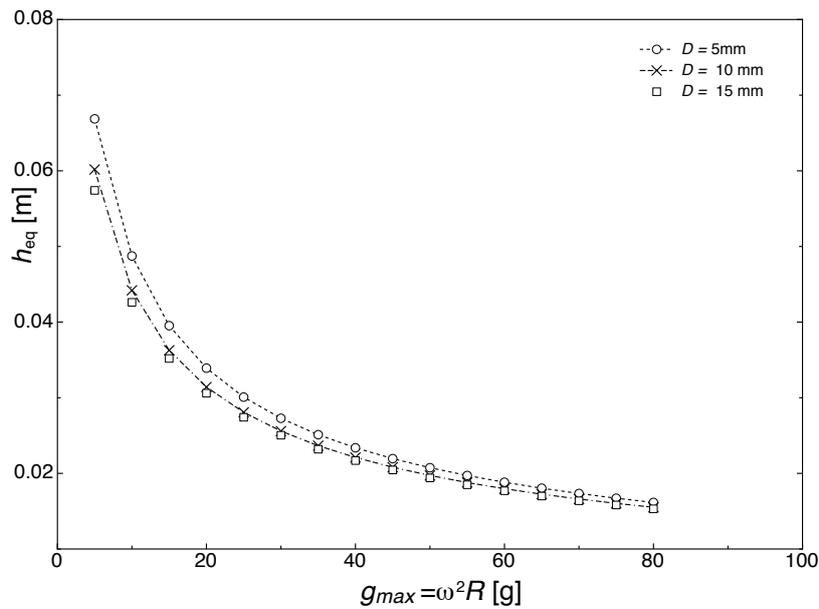

**Fig. 9** Simulated height data for a critical strain of 0.02 for three tube diameters.

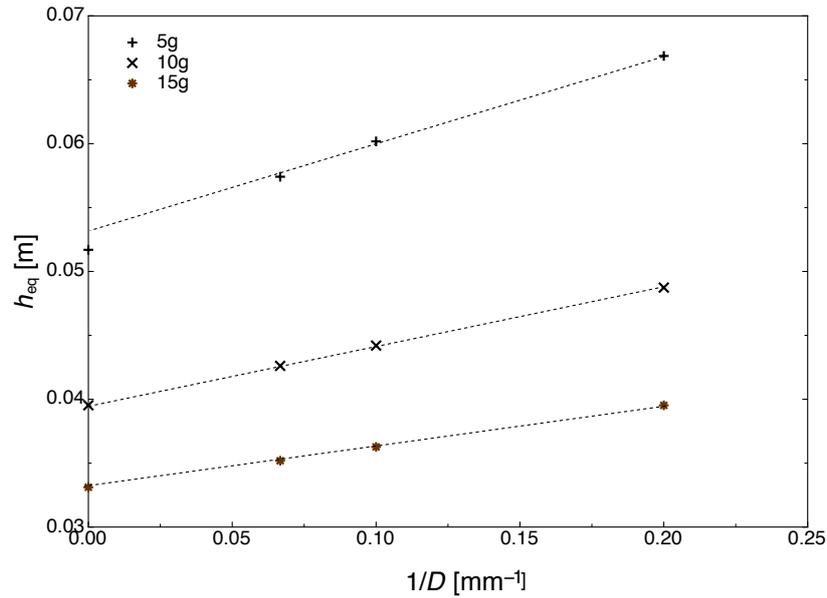

**Fig. 10** Extrapolation of simulated height data for a critical strain of 0.02 to infinite tube diameter for three accelerations measured in units of normal gravity g. The dashed lines are linear extrapolations of the three data points for finite $h_0$. Here though, the points for infinite $D$ are known. At the lowest g, near the gel-point, the overall trend is somewaht convex to the abscissa implying that several more tube diameters would be needed in practice in order to obtain a good extrapolation.

**Concluding remarks**

The approximate equations used conventionally to calculate the compressive strength curve for cohesive particulate gels from raw centrifuge data ignore the possibility of adhesion to the walls. Furthermore, they are based on the assumption that the acceleration vector field inside the tube is everywhere parallel to the tube axis. The validity of these two assumptions has been explored and tested using simulated test data, as has the validity or otherwise of some possible correction factors for adhesion suggested by dimensional analysis. Corrections to the "parallel g" assumption are however very unlikely to ever be needed in practice, as this approximation turns out produce negligible error for realistic centrifuge geometries. Furthermore, it turns out that for realistic or typical centrifuge geometries, by which is meant $h_0/R \ll 1$ and $D/R \ll 1$, it is simpler and more accurate to use eqns 6 (but with $g$ replaced by the mean centrifugal acceleration) than it is to use the more complicated but still approximate eqns 7. The reasons for this are two-fold: the first being that the extra terms in eqn 7 are insignificant for sufficiently small $h_0/R$ and $D/R$, the second is that replacing $g_{max}$ by $g_{mean}$ improves the accuracy of eqn 7.

Adhesion was parameterised in terms of a critical shear strain for shear yielding at the wall. This was done simply for convenience, as a way of parameterising the intrinsic strength of adhesion for any system overall. This parameter is useful because the ratio of the shear yield stress to compressive strength itself is expected to vary strongly with volume-fraction near the gel-point (before becoming more or less constant at high volume-fraction), whereas the apparent critical shear strain is known to vary much more weakly with concentration [34]. The apparent critical shear strain has been found to vary from one material to another and with particle size. Hence, here, values of 0, 0.0002, 0.002 and 0.02 were used in order to cover the range reported in the literature, wherein values of << 0.01 have been reported for many electrolyte coagulated or flocculated suspensions, including polystyrene latex and a range of minerals, whereas values of order 0.02 have been reported for one electrolyte flocculated mineral system [27]. The latter work apart, critical strain values of order 0.01 or more have only been reported otherwise for mineral suspensions strongly-flocculated using high polymers and for water-treatment sludges[27]. A particular form for the dependence of the shear and compressive strength on volume-fraction had to be assumed and this again was motivated by published experimental data[27]. The dependence of the resulting adhesive to compressive strength ratio on volume-fraction and critical strain was plotted in Fig 1 in order to illustrate the parameter space covered overall.

Small adhesive to strength ratios alone do not necessarily mean a negligible effect of adhesion in the centrifuge, since the adhesive effects are amplified by the geometric factor $\sim 4h_0/D$ and this factor tends to be rather large and in the range $10 < 4h_0/D < 100$ for off-the-shelf centrifuge rotors and tubes.

The simulations herein suggest that adhesion can probably be neglected in centrifuge testing of many if not most systems, for volume-fractions well above the gel-point, at least, by which is meant $>> 2\phi_g$, and otherwise where the apparent critical shear strain $<< 0.01$. More generally, the following tactics might be considered:

1. Measure the adhesive shear strength independently in a rheometer fitted with appropriate tools and apply a correction to (6) or (7) using (8) and (10). The simple corrections suggested here will always fail very near the gel-point though, simply because adhesion dominates there.

2. Make measurements using several different values of the ratio of initial height $h_0$ to tube diameter $D$ and extrapolate to $h_0/D$. This method may or may not be practicable, depending upon the design of the centrifuge. The range of $h_0/D$ covered should be at least three-fold and preferably five-fold, as should the number of points, although it rather depends upon the strength of the effect. Fewer points would needed to demonstrate that adhesion is insignificant, where it is, than would be needed to perform a good extrapolation, where it is not.

3. Complement the centrifuge data with data from gravity-settling for a set of column chosen so as to vary $h_0/D$ by a factor of at least five and better ten, with $h_0$ and $D$ each varied by at least three. One could also complement centrifuge data with that coming from an instrumented pressure-filter, in principle, since the construction of these is usually such that $h_0/D$ is small, except that pressure-filtration does not normally work near the gel-point because of the difficulty in applying the necessary small pressures. This method has however been used to validate and complement the centrifuge method at higher volume-fraction [8, 10, 21, 23].

The centrifuge method has been used recently to examine the effect of temperature on the stability of incipiently flocculated polystyrene latices [37], but without taking the possibility of adhesion into account. It has been used also to measure the compressibility of microgels [38], albeit using inclined rather than horizontal or swing-out rotors which adds complication to the analysis. Wall shear stresses were ignored, which could well be justifiable for microgels in good solvents, perhaps [38].

A concise discussion of the essential constitutive behaviour of strong particulate gels can be found elsewhere [34]. In that work it was shown that if the gel-point is known from some independent measurement, of, for example, G*, say, the complex shear modulus, or from sedimentation tests in squat tubes at 1g, then the actual ratio of adhesive to cohesive shear yield strengths, α, say, can be found from data analogous to that shown in fig. 8 by correcting the apparent compressional data using guesses for α and shooting for the known gel point. By such means, a value of ca. α = 1/6 was deduced for coagulated suspensions of AKP-30 alumina in polycarbonate tubes[34].

**Table 1: Values of key quantities used in the simulations**

| Quantity | Symbol | Value |
|---|---|---|
| Initial vol. fraction | $\phi_0 = \phi_g$ | 0.10 |
| Compressive strength Pre-factor | $k$ | 5 Pa |
| Compressive strength exponent | $n$ | 4, 5 |
| Interphase density difference | $\Delta\rho$ | 2000 kg m$^{-3}$ |
| Tube diameter | $D$ | 5, 10, 15 mm |
| Initial column height | $h_0$ | 0.075m |
| Centrifuge radius | $R$ | 0.175m |

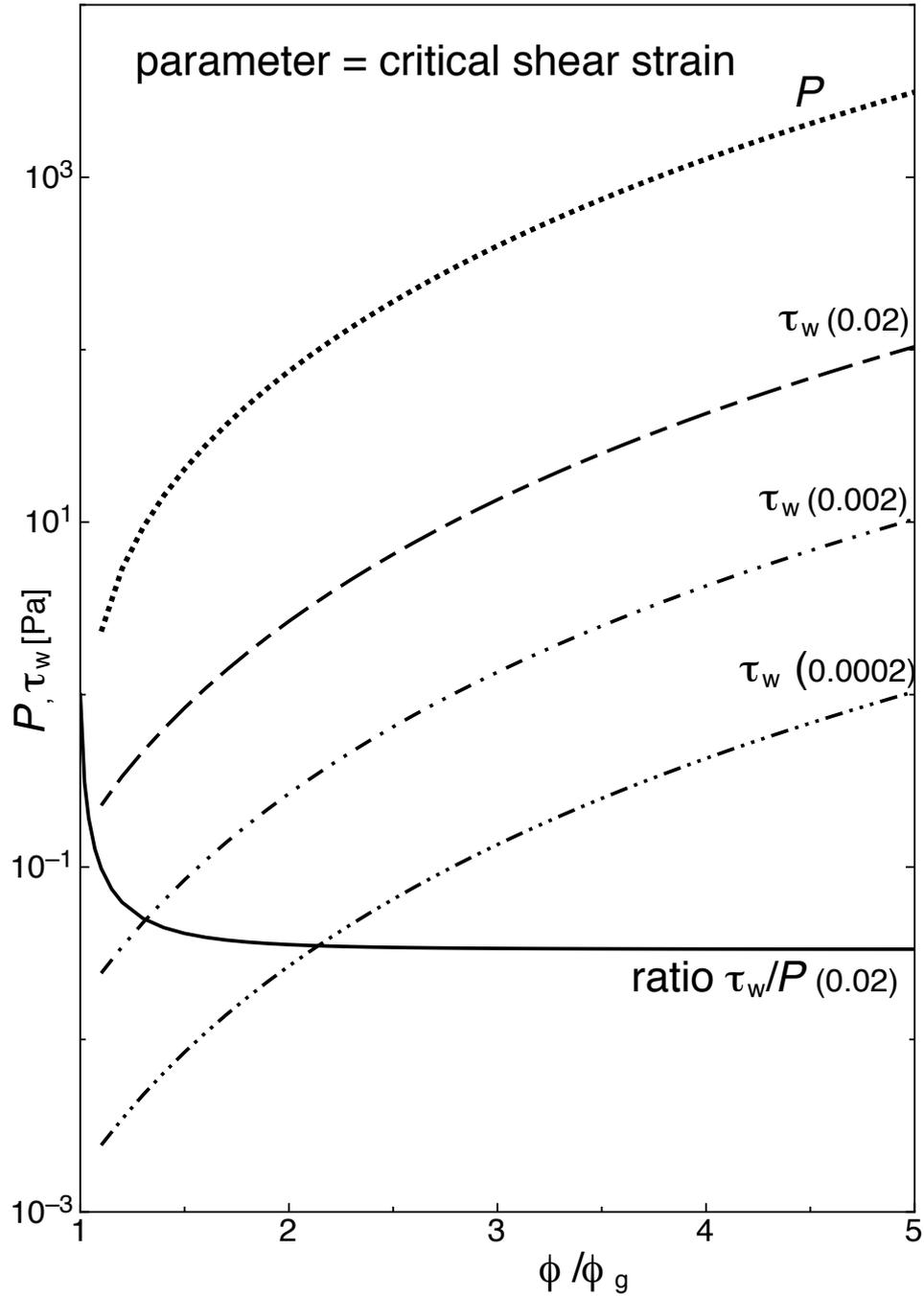

**Fig. 1** Input material functions: compressive strength and wall stress, the latter for three values of the apparent critical shear strain. The ratio of wall stress to compressive stress is plotted for the largest strain. It can be seen that most of the variation is confined to a region close to the gel-point.

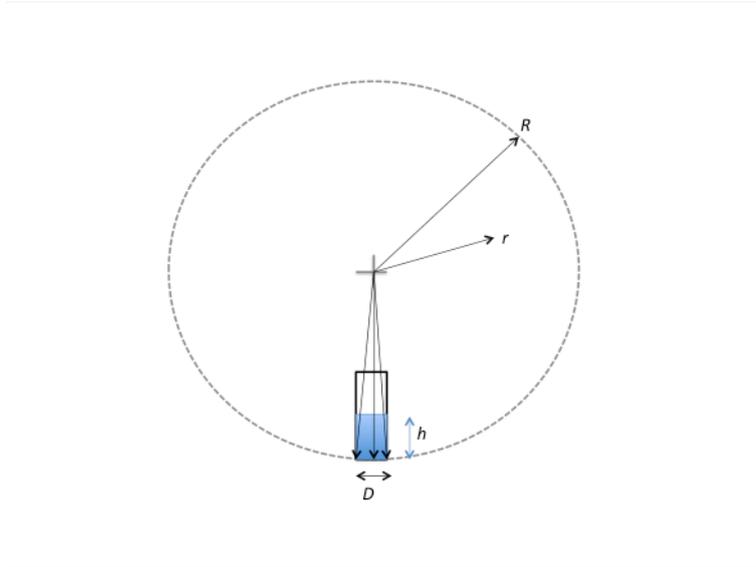

**Fig. 2** Schematic illustration of the geometry of the centrifuge test.

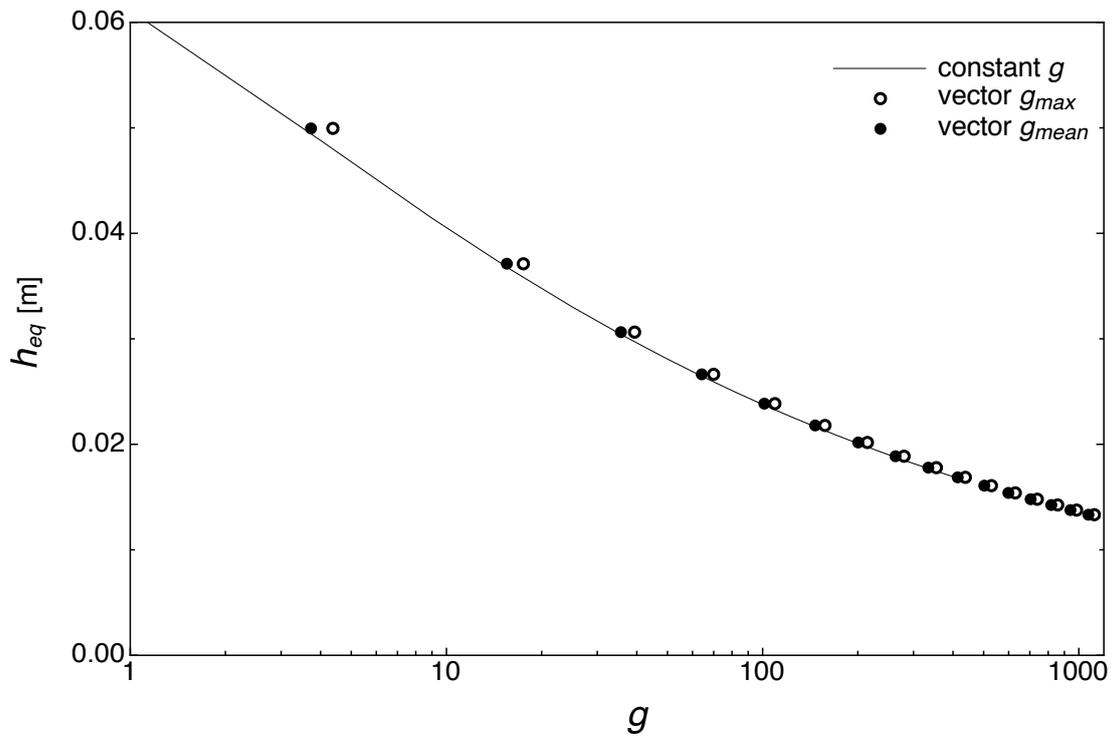

**Fig. 3** Synthesised equilibrium height versus acceleration data for constant $g$ compared with centrifugal acceleration at equivalent mean acceleration $w^2(R-h/2)=g$ for $n=4$, $D=15$mm and $h_{eq}/R < 1/3$.

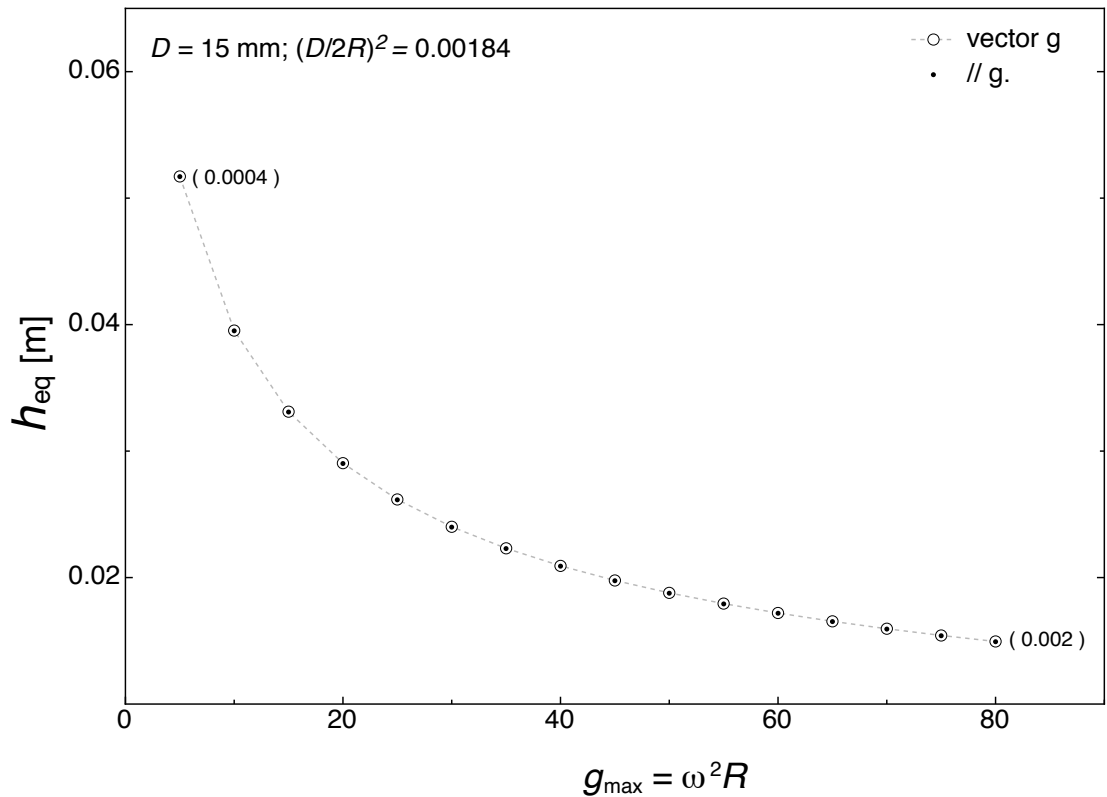

**Fig. 4** Synthesised equilibrium height versus acceleration data for vector centrifugal $g$ compared with the parallel-$g$ for 15mm internal dia. tubes and $n= 4$. The figures in brackets show the fractional error of the parallel-g approximation at each end of the curve.

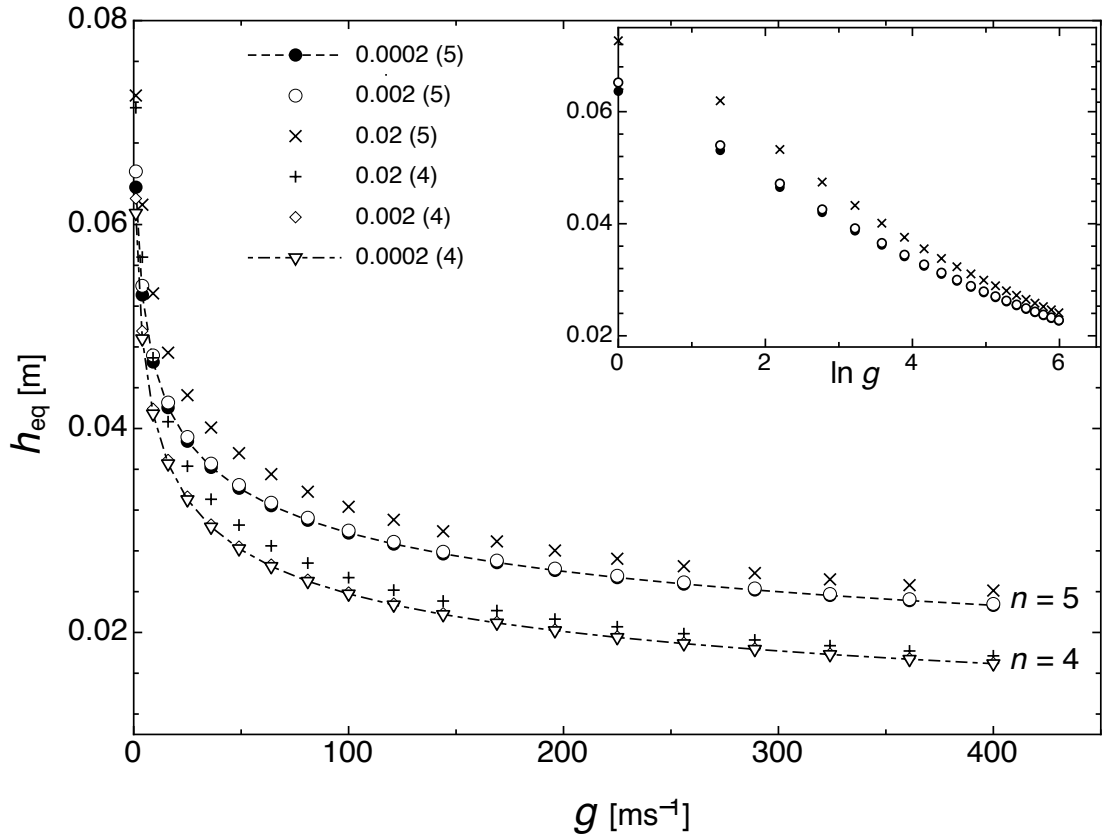

**Fig. 5** Synthesised equilibrium height versus acceleration data for constant $g$, 10mm internal diameter tubes, two values of the power-law index and for three different levels of wall adhesion as parameterised by the critical shear strain. The effect of adhesion is just discernible at 0.002 and significant at 0.02. The result for zero shear strain (not shown) is indistinguishable from the curve for 0.0002. The inset semi-log plot illustrates the slow change in the gradient term needed in eqns (7) and (6).

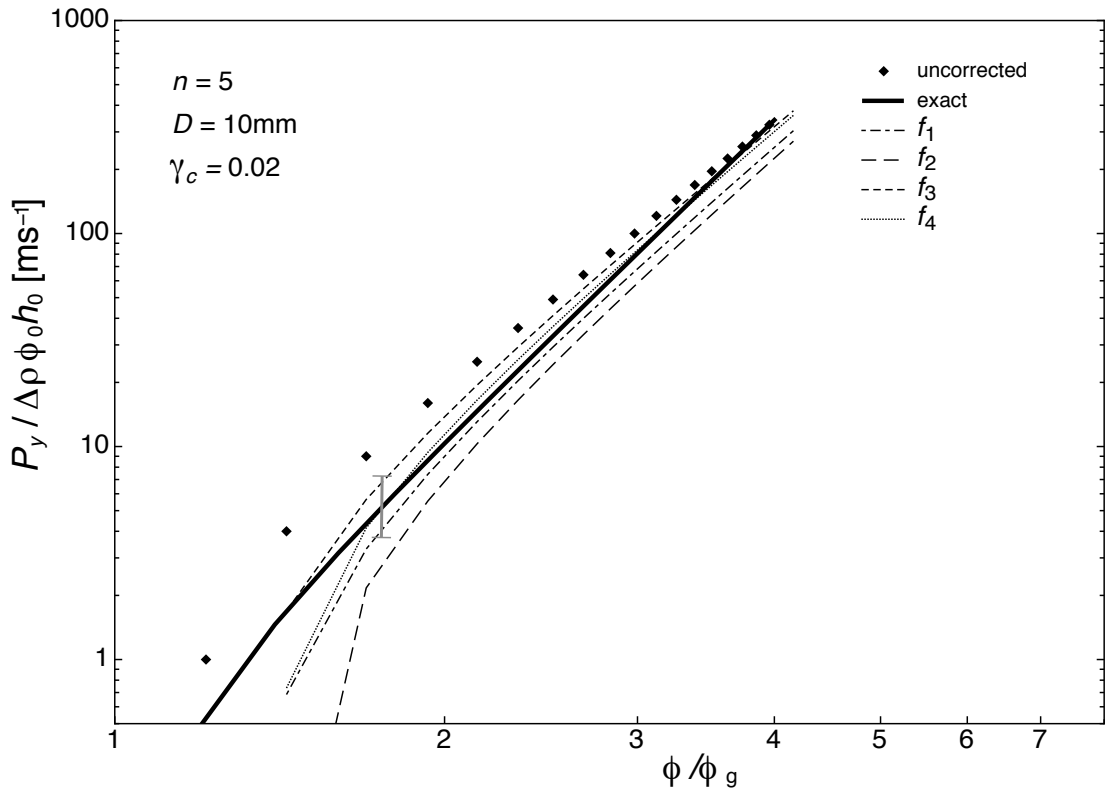

**Fig. 6** Apparent compressive strength for critical adhesive shear strain = 0.02 obtained by applying eqns 3 to the data in fig. 5 (points) compared with the true curve (heavy line). Also shown are the results of applying the suggested under and over corrections $f_1$ to $f_4$ to the points. Approximate corrections of this type cause the result to become negative below ca. 1.3 times the gel-point. Below that only iteration of the data via retrodiction and correction will do and even that scheme will fail if the gel is supported entirely by the walls.

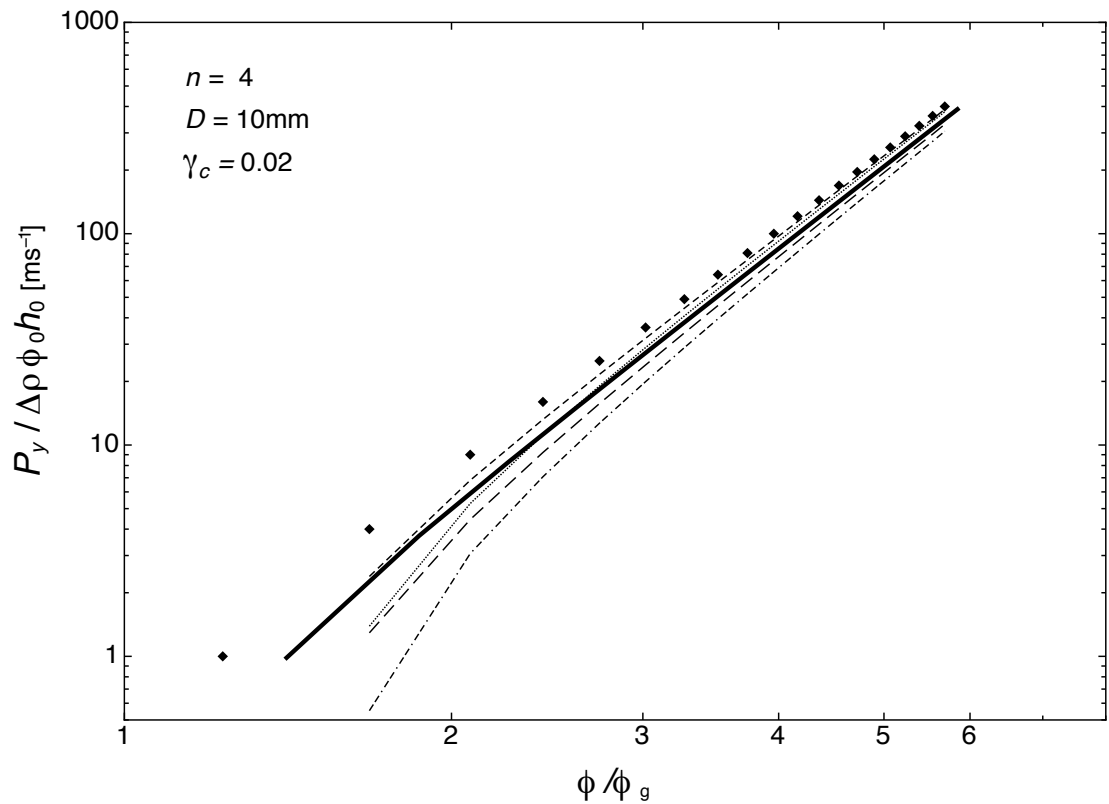

**Fig. 7** As for fig. 6 but for a power-law index of $n = 4$. Reducing the power-law from 5 to 4 reduces the errors coming from adhesion a little, but otherwise the picture is much the same.

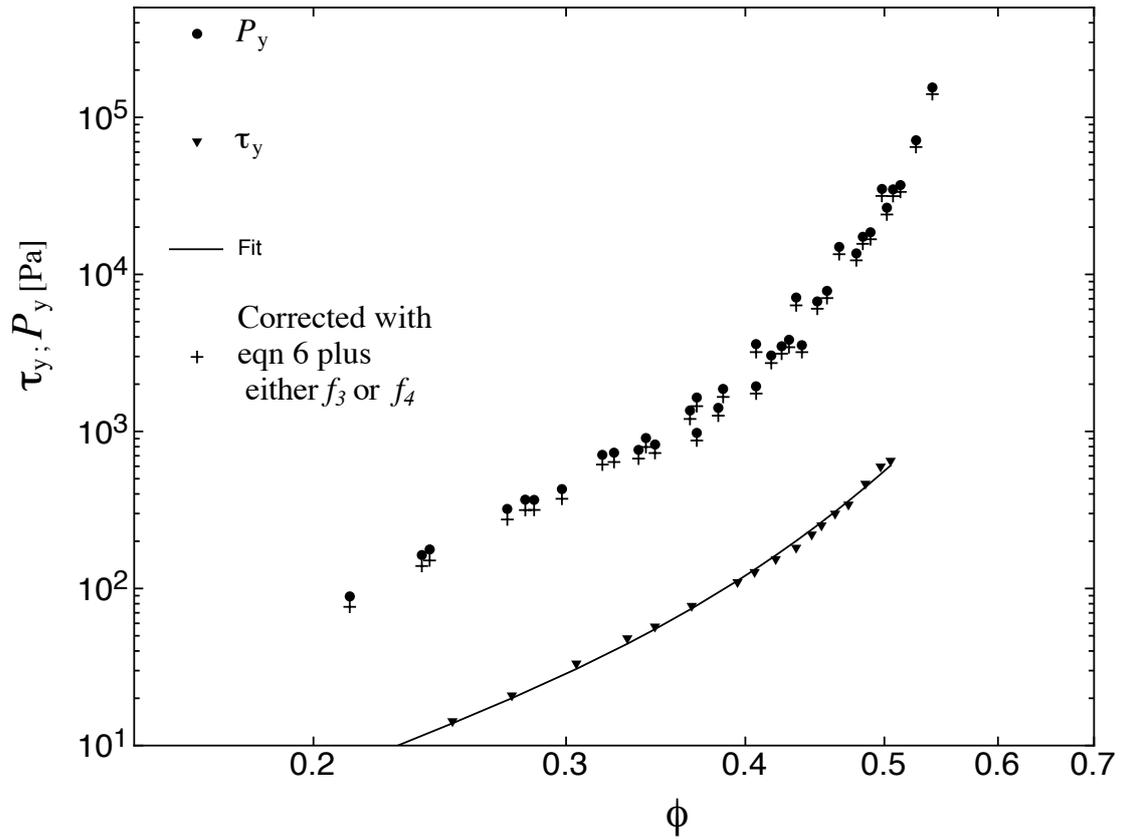

**Fig. 8** Data of Zhou et al. (replotted from ref. 10, see also ref. 8, fig. 3) and corrected according to eqn 6 using correction $f_3$ from eqn 10 and the fit to the shear data shown. Note that correction $f_4$ gives very similar results.

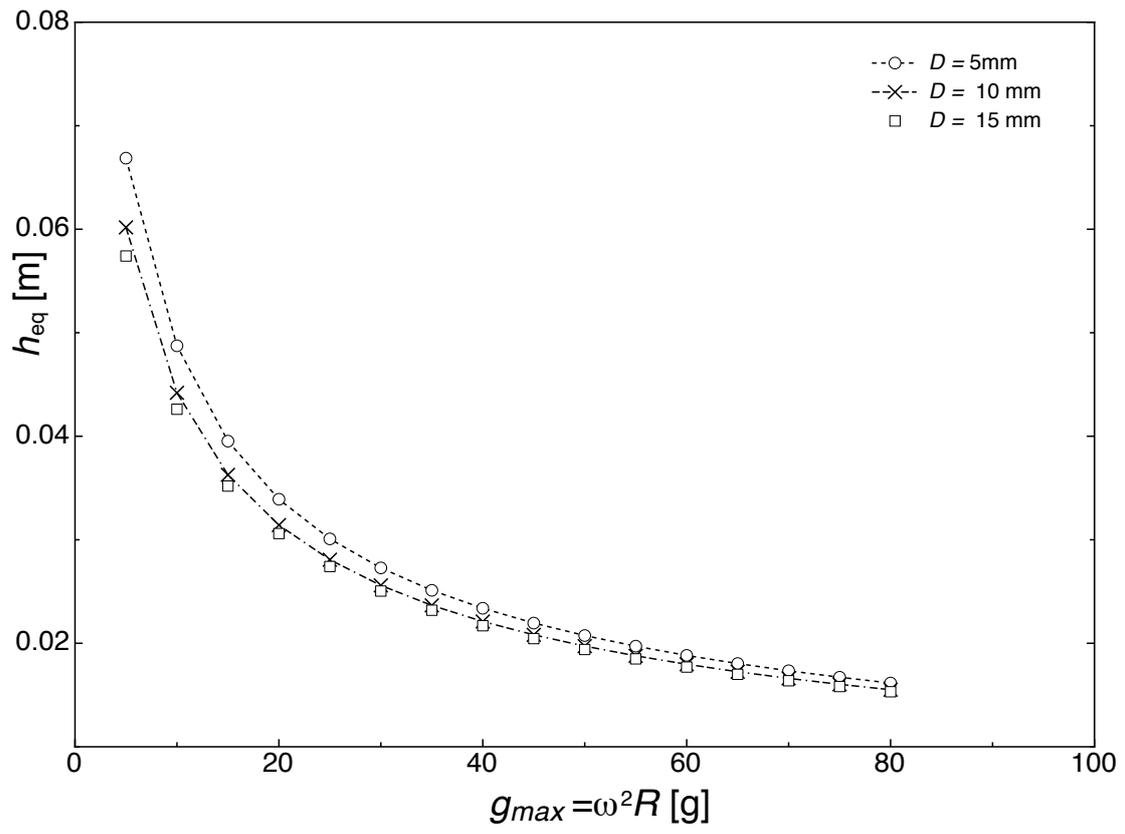

**Fig. 9** Simulated height data for a critical strain of 0.02 for three tube diameters.

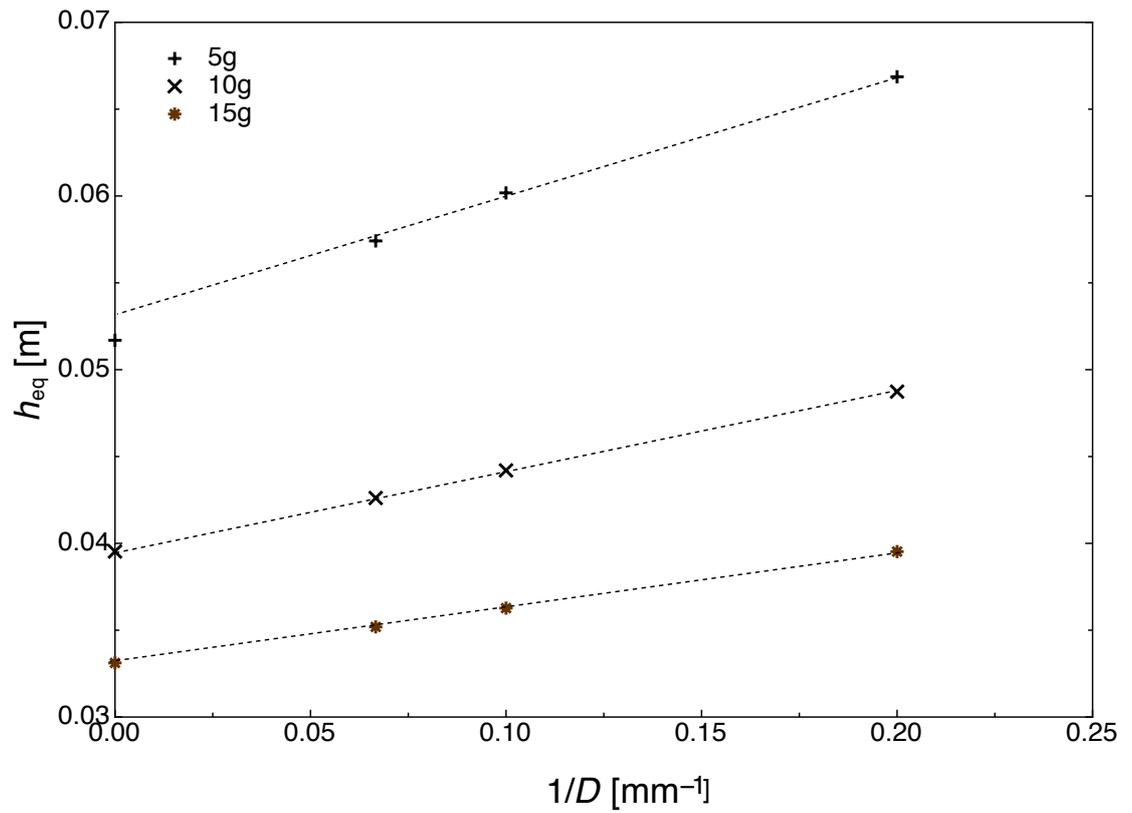

**Fig. 10** Extrapolation of simulated height data for a critical strain of 0.02 to infinite tube diameter for three accelerations measured in units of normal gravity g. Here the points for infinite *D* are known. At the lowest g, near the gel-point, the plot is slightly convex to the abscissa implying that several tube diameters would be needed in practice for a really good extrapolation.